\documentclass[aps,prl,twocolumn,showpacs,floatfix]{revtex4}

\usepackage{subfigure}
\usepackage{graphicx}
\usepackage{dcolumn}
\usepackage{epsfig}
\usepackage{amssymb}
\usepackage{amsmath}
\usepackage{rotating}
\usepackage{color}

\begin{document}
\title{Slippage of water past superhydrophobic carbon nanotube forests in microchannels}

\author{P. Joseph$^{1}$, C. Cottin-Bizonne$^{1}$, J.-M. Beno\^it$^{1}$, C. Ybert$^{1}$, C. Journet$^{1}$, P. Tabeling$^{2}$, L. Bocquet$^{1}$}
\email{lyderic.bocquet@univ-lyon1.fr}
 \affiliation{(1) Laboratoire PMCN,
Universit\'e Lyon 1, CNRS, UMR 5586
69622 Villeurbanne, France \\
 (2) Physico-Chimie Th\'eorique, ESPCI,  CNRS, UMR 7083   75231 Paris, France\\}

\begin{abstract}
We present in this letter an experimental characterization of
liquid flow slippage over superhydrophobic surfaces made of carbon nanotube forests,
 incorporated in
microchannels. We make use of a $\mu$-PIV (Particule Image Velocimetry)
technique to achieve the submicrometric resolution on the flow
profile necessary for accurate measurement of the surface
hydrodynamic properties. We demonstrate boundary slippage on the
Cassie superhydrophobic state,
associated with slip lengths of a few microns, while a vanishing slip length is
found in the Wenzel state, when the liquid  impregnates the surface.
Varying the lateral roughness
scale $L$ of our carbon nanotube forest-based superhydrophobic
surfaces, we demonstrate that the slip length varies linearly with
$L$ in line with theoretical predictions for slippage on patterned surfaces.
\end{abstract}
\maketitle
%
%
The nature of the hydrodynamic boundary
condition of a liquid past a solid surface has been actively
revisited over the recent years, strongly motivated by the
emergence of microfluidics and nanofluidics \cite{ref,Squ05}.
Downsizing leads to large surface-to-volume ratio, and the flow
properties at the solid boundaries become a key factor in the
understanding of the motion of liquids in ever smaller channels
\cite{Squ05}. In this context the violation of the no-slip
boundary condition (BC) and the existence of slippage of the fluid
at the solid surface is a promising way to bypass the huge
increase in hydrodynamic resistance that comes with
miniaturization.
\par
Slippage, usually quantified by a slip length $b$, has been
demonstrated on solvophobic surfaces, with a slip length of a few
tens of nanometers \cite{BB,ref,Cot05,Joly06,Vino03}. Such values are
obviously insufficient to modify the flow in channels with
micrometer sizes and other specific solutions have to be proposed
to benefit from slippage effects in microfluidics.
Super-hydrophobic surfaces are promising in this context, since it
has been recently predicted \cite{Cot} and experimentally reported
\cite{Roth05} that the associated ``Fakir'' effect (the
so-called Cassie state) considerably amplifies slippage. This
situation is achieved with highly rough hydrophobic surfaces:
instead of entering this bidimensionnal hydrophobic porous
medium, the liquid remains at the top of the roughness thus
trapping some air in the interstices, and therefore leading to a
very small liquid-solid contact area.
\par
In the context of microfluidics, the challenge now consists in
developping versatile methods to design such surfaces in
microchannels with optimized flow properties. These have to deal
with two conflicting constraints on the engineered surfaces: low
friction --{\it i.e.} large slippage-- and robustness of the Fakir
effect against pressure induced impregnation. On the one hand,
large slippage is achieved by minimizing
the liquid-solid area together with maximizing the underlying
lateral length scale $L$ of the roughness. Indeed, theoretical
predictions show that the resulting effective slip length $b_\mathrm{eff}$ is
mainly fixed by the roughness scale $L$
\cite{Stone,Philip,Cottheory,cyb06}: $b_\mathrm{eff}\simeq \alpha L$, with
$\alpha$ a numerical factor (smaller than unity for typical
surfaces). On the other hand, the stability of the Fakir (Cassie)
state with respect to pressure variations fixes some restrictive conditions
on the underlying roughness scale $L$ \cite{quere2002}: above an
excess pressure of the order of $\Delta P_{int}\sim 2\gamma_{LV}
/L$ (with $\gamma_{LV}$ the liquid-vapor surface tension),
penetration of the liquid into the roughness occurs --Wenzel
state--, and the benefits from the low friction Fakir state are
lost. The potential applicability of lithographically designed
patterned surfaces, with a surface lateral scale $L$ in the ten
microns range is therefore limited by their low
resistance to pressure \cite{Roth05}. While one would expect the
increased robustness on nanoengineered surfaces to be achieved at
the expense of their slippage properties, recent rheology
measurements on nanopost superhydrophobic surfaces have reported
slip lengths of tens of microns \cite{Choi} in clear contrast with
theoretical expectations \cite{Stone,Philip,Cottheory,cyb06} 
 and with the
experimental results of \cite{Roth05} at larger (micrometric) scales. 
This surprising result would suggest that the underlying roughness 
does not limit the slip length for nanopatterned surfaces, as first expected.
However the sensitivity of the rheological method used in \cite{Choi}
appears questionnable \cite{Boc06}, and it therefore points out
the need for accurate experimental data to achieve a
proper understanding of the role of the underlying roughness
scale, from nano- to micro- patterned surfaces.
\par
The purpose of the present work is precisely to provide accurate
data allowing 
to understand the role of the roughness scale on the
hydrodynamic properties of superhydrophobic surfaces.
Taking
benefit of recent advances in nanomaterials science we
incorporated versatile composite surfaces in microchanels. We make
use of recently developed surfaces covered with vertically
aligned and densely spaced carbon nanotubes (CNT
forests)~\cite{Lau2003,EPL05}. Together with the existence of
large contact angles (larger than $165^\circ$), the submicrometric
lateral length scale achievable on these surfaces was shown to
provide a robust Fakir state with respect to pressure \cite{EPL05}, with an
intrusion pressure $P_{int}\sim 1{\rm atm}$. More
interesting in the present framework, is the ability provided by
the method to achieve various lateral length scales $L$ in the
submicron to microns range. Flows in the vicinity of such surfaces
were then investigated using particle image velocimetry at the
micron scale ($\mu$-PIV) to probe the velocity profile in the
microchannels with submicrometric resolution. Our measurements
confirm that a low friction is achieved providing the interface
lies in the Fakir state, and demonstrate that these low friction
properties are indeed controlled by the lateral roughness scale,
in agreement with theoretical considerations \cite{Cottheory,
cyb06}.
\par
%
\begin{figure}[t]
\centering
\vspace{-.5cm}
{\includegraphics[width=8cm]{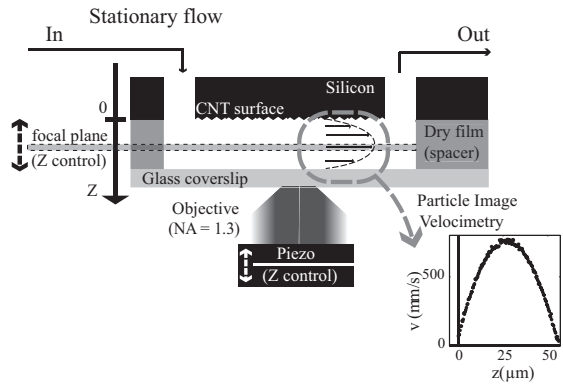}}
\caption{\label{fig1} Experimental setup. A pressure driven flow
is conducted in a glass/NTC micro-channel. The focal plane $Z$ is
controlled with a piezo, a large numerical aperture objective
allows for a narrow depth of field. Velocity is measured by
particle image velocimetry over
the whole channel depth.
}
\vspace{-.7cm}
\end{figure}
\begin{figure}[t]
\begin{center}
{\includegraphics[width=7cm]{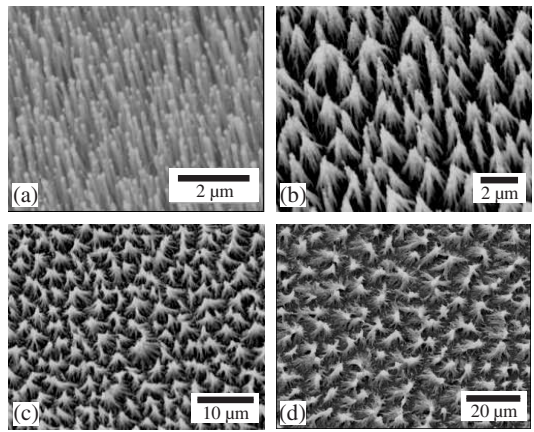}}
\end{center}
\caption{\label{fig2} SEM pictures of superhydrophobic CNT
forests, after  functionalization with thiols: (a) in the gas
phase; (b-d) in ethanol, for increasing initial nanotube length,
resulting in characteristic lateral roughness length scale $L$:
1.7, 3.5 and 6~$\mu$m (pictures b, c and d).}
\vspace{-.7cm}
\end{figure}

The experimental setup is shown in fig. \ref{fig1}. The CNT
forests are obtained using Plasma Enhanced Chemical Vapor
Deposition (PECVD), using the protocol previously described in
\cite{Lau2003,EPL05}. The CNT forests are grown on a silicon
substrate covered with a SiO$_{2}$ layer and a thin film of Ni
catalyst. This develops Ni islands through the sintering of the
Ni film (after introduction in a  furnace heated to 750$^{\circ}$C
under 20 torr of hydrogen). Ammonia and then acethylene
(C$_{2}$H$_{2}$) are introduced into the furnace to initiate the
CNT growth. A dc discharge between the cathode and anode samples
allows for directed growth.
The depositions are carried out between 5 and 60
minutes in a stable discharge. Typical CNT forests grown through
this process are characterised by a typical nanotube diameter
between 50 and 100 nm, a nanotube inter-distance 100-250 nm, and
micrometric lengths, as shown on Fig. \ref{fig2}(a). The nanotube
lengths and inter-distance can be tuned by varying the deposition
time and the thickness of the nickel layer.
\par
To get a superhydrophobic behavior, the CNT forests are then
functionalized using thiols in gas or liquid phase, after covering
them with an anchoring thin gold layer (5 nm) by sputtering as
described in \cite{EPL05}. While the gas phase functionalization
leaves the CNT forest structure unchanged \cite{EPL05}, the
functionalization in a liquid phase results in bundles of carbon
nanotubes (see fig~\ref{fig2}(b-d)), as a result from capillary
adhesion during the evaporative drying of the functionalization
solvent, ethanol \cite{Bico2004,EPL05}. Combined with the control
of the nanotube density and length, this bundling process is used
in the present study as a further way to tune the roughness
lateral scale $L$.
%
\par
These CNT surfaces are embedded in a microfluidic setup (see
fig.~\ref{fig1}) featuring a microchannel  of height 40-$50 \mu$m,
width $2$mm and length $2$cm. This microchannel is obtained by
sandwiching a dry film resist that has a lithographic trench
between a CNT covered silicon substrate and a microscope glass
coverslip, following a recently introduced method~\cite{Vulto05}.
A pressure driven flow is set up by controlling inlet and outlet
pressures in the range of -50 to 100~mbar with respect to the atmospheric
pressure, while the pressure drop along the channel remains below
2~mbar. The working solution is deionized water seeded with
fluorescent particles (500 nm in diameter), $10^{-4}$ in
volumetric concentration. The observation is made with a LEICA
epifluorescent microscope, using a $100\times$ oil immersion large
numerical aperture objective (NA=1.3) achieving a 700nm depth of
field. An optical depth of 500nm is obtained for the measurement
volume by thresholding on the intensity and then selecting the
particles in focus. The velocity profile is measured using a
standard $\mu$-PIV technique, similar to that described in
\cite{Joseph2005}, where further details can be found. Briefly, a
scan in the vertical $z$ direction is provided using a
piezotransducer with $0.2\mu$m step increments, allowing to
measure the velocity profile $v(z)$ in the whole microchannel. A
time-delayed intensity cross correlation is performed in a
$20\mu$m$\times 60\mu$m window, and the computed velocity is
averaged over 20 pairs of images for each $z$-position of the
piezo. We thus obtain eventually an averaged velocity in an imaged
zone of $500 {\rm nm}\times 20 \mu{\rm m}\times 60\mu{\rm m}$.
Typical velocity profiles (see fig. \ref{fig1}) present the
characteristic parabolic shape associated with the pressure-driven
Poiseuille flow.
\par
\begin{figure}[t]
\centering
{\includegraphics[width=6.5cm]{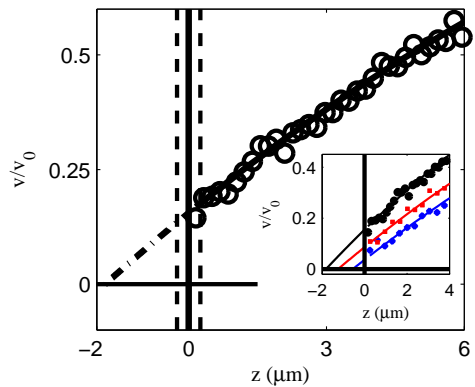}}
\vspace{-.6cm}
\caption{\label{fig3} Zoom of the velocity profile in the region
close to the CNT surfaces (in the Cassie state), as measured by the
$\mu$-PIV method ($\circ$), normalized by the velocity in the
center of the channel $v_0$ ($v_0=360 \mu{\rm m/s}$). The errorbar on 
the velocity is $\sim 10\mu m.s^{-1}$, of the order of the size of the
symbols (see \cite{Joseph2005}).
The two vertical dashed lines
correspond to the extrema of the measured position of the CNT
surface, while the vertical solid line is the averaged position of
the interface. The dashed-dotted line is the fitted velocity profile,
using the parabolic (Poiseuille) prediction and the Navier
BC. Inset: the three different profiles correspond
to CNT surfaces with different roughness length scale $L$. From
bottom to top, $L= 1.7,\ 3.5,\ 6\ \mu$m, and
$v_0 = 550; 760; 360 \mu{\rm m/s}$. 
Slippage is shown to increase with $L$.  \\
}
\vspace{-.7cm}
\end{figure}

The position of the CNT surface has been carefully determined. As
in \cite{Joseph2005}, we take advantage of the presence of a few
particles which unavoidably adsorb on surfaces [we only considered
windows of observation where at least 3, and up to ten, of such
beads are included]. The position of the surface is then
determined by fitting the fluorescence intensity emitted by these
beads with a lorentzian and by substracting the bead radius from
the location of the maximum of the lorentzian. We have considered
flows on surfaces in the Cassie state (with trapped air), but also
on surfaces in the Wenzel state, where the CNT surfaces are
impregnated. This latter behavior --impregnation of the surface
roughness-- occurs for a few devices at long times and is likely
due to defects at the edges of the channel (as suggested by direct
observation). These few devices therefore offer the opportunity to
also study slippage in the Wenzel state. When the CNT surface is
in the Cassie state, the liquid interface lies at the top of the
nanotube bundles, to which adsorbed beads are therefore
restricted, fixing thereby the effective solid-liquid interface.
On the other hand, in the Wenzel state beads are distributed
between the top and bottom of the nanotubes and the ``effective"
(solid-liquid) interface is in that case fixed at the mean
position between the top and the bottom of the CNT. The CNTs
present a length polydispersity, responsible for the uncertainty
in the location of the liquid interface. This uncertainty is
measured on each CNT surface and is typically of the order of $\pm
200$nm, similar to the characteristic deviation in height of the
grown CNT forests (see figs \ref{fig2} and \ref{fig3}).
\par
Fig.~\ref{fig3} shows a
zoom of the water velocity profile measured in the vicinity of the
liquid interface when the CNT surface is in the Cassie state. The
two vertical dashed lines represent the uncertainty in the
position of the liquid interface, as discussed above. As shown in
this figure, the velocity profile does not vanish at the boundary and does exhibit
slippage at the interface. This slippage can be accounted for by the
standard Navier BC for the velocity field at the
interface $b\,\partial_z v= v$: $b$ is the slip length and can be
interpreted as the extrapolated distance at which the velocity
would vanish. Experimentally, $b$ is obtained by fitting the
velocity profile using the Poiseuille prediction.  For each
surface, this procedure is repeated on about 5 measurement windows
chosen in different locations at the CNT surface in the
microchannel from which an averaged value and error bar are
deduced for the slip length.
%
\par
\begin{figure}[t!]
\centering
{\includegraphics[width=7cm]{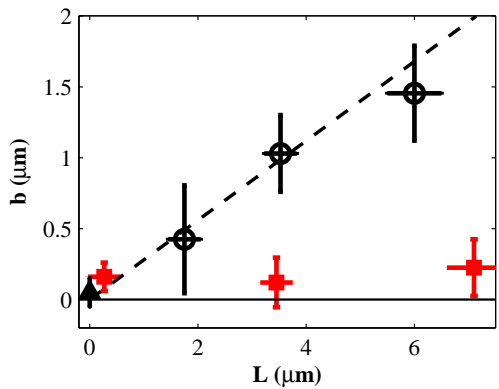}}
\vspace{-.5cm}
\caption{\label{fig4} Measured slip length on the CNT surfaces as
a function of the characteristic lateral length scale $L$ of the
underlying CNT forest. $\circ$ : surfaces in the Cassie state;
$\blacksquare$ : surfaces in the Wenzel state; $\blacktriangle$ :  benchmark measurement on
the bare silicium (without CNT forest). The dotted line
is the theoretical prediction for the slip length, $b_\mathrm{eff}=\alpha(\phi_S) L$,
with $\alpha=0.28$ (computed from the best linear fit).}
\vspace{-.8cm}
\end{figure}

As mentioned above, the characteristics of the underlying CNT
surfaces can be varied: the distance between bundles is mainly
tuned by the length and density of the forest before
functionalization in the liquid phase \cite{EPL05}. 
We have examined the flow behavior on surfaces with
various lateral length scales $L$, measured by computing the intensity 
autocorrelation of the top images of the SEM pictures. In the examined 
surfaces this typical lateral characteristic length of
the roughness varied between $1\mu$m up to $6 \mu$m. 
We have therefore measured the slip length on these
surfaces using the procedure described above. Resulting flow
profiles for the Cassie state are shown in the inset of fig.
\ref{fig3}. The slippage increases with the
roughness length scale $L$ as is summarized in fig. \ref{fig4}
where the measured average slip length $b$ is plotted against $L$,
the error bar corresponds to the dispersion in the measurements. 
We have also measured the slip length for different $L$ for
systems in the Wenzel state. 
As shown in fig. \ref{fig4},
slippage is strongly reduced in the Wenzel state, due to the increased
liquid-solid friction in the absence of trapped air in the CNT, in agreement
with theoretical predictions \cite{Cottheory}.
%
\par
Our results point out the absence of slippage
in the Wenzel sate and the direct, linear, dependence of the
measured slip length on the CNT lateral roughness scale $L$ in the
Cassie superhydrophobic state. This confirms recent theoretical
predictions made for the effective slip length on superhydrophobic
micro-structured surfaces \cite{Philip,Stone,Cottheory,cyb06}. In
the Cassie state, the liquid is in direct contact with the solid
only through a small fraction ($\phi_S$) of the surface. At the
hydrodynamic level, this composite surface leads to a spatially
dependent BC with a no-slip BC on the fraction of
real solid-liquid interface (with fraction $\phi_S$), while the
remaining surface is characterized by a perfect slip BC 
($b=\infty$) on the suspended liquid-vapor interface. At
a qualitative level, the theoretical results can be summarized by
stating that the effective slip length $b_\mathrm{eff}$ essentially
saturates at the value fixed by the lateral scale of the roughness
$L$, $b_\mathrm{eff}=\alpha(\phi_S) L$. For instance a flow perpendicular
to stripes yields $\alpha=\frac{1}{2\pi} \log(1/\cos[\frac{\pi}{2}(1-\phi_S)])$ \cite{Philip,Stone} showing a weak dependency on
$\phi_S$ for feasible surfaces.
In our case, the experimental surface is closer to an array of
pillars. Since no analytical form is
available in this case, we have extended our previous theoretical analysis
of Ref. \cite{Cottheory} to this geometry,
showing that the main features of the
effects described in \cite{Philip,Stone,Cottheory} (for a surface made of stripes)  are preserved
 for a pattern of pillars \cite{cyb06}. In particular,
our predictions in this geometry are again summarized by
$b_\mathrm{eff}=\alpha(\phi_S) L$, with a slowly varying
$\alpha(\phi_S)$, except very close to $\phi_S=0$. This prediction
is shown in fig. \ref{fig4} to be in good agreement with
experimental results, with a value of $\alpha\simeq 0.28$, corresponding
to a constant value for $\phi_S$ of 0.15--0.2. The latter is reasonably consistent with the
experimental value of $\phi_S\approx 0.1$ estimated from the
contact angle on the CNT surfaces ($\approx 165^\circ$) and the
Cassie theory  \cite{EPL05}. The slightly higher value of $\phi_S$
obtained in the slippage measurements may result from the random
distribution of pillars, as opposed to the  regular one used here
in the theoretical description of slippage.
\par
Finally the slip lengths we measure are substantially smaller
than the ones reported recently in \cite{Choi} using global
rheological measurements ($b_\mathrm{eff}\sim 20\mu$m for water).
However, unlike the present $\mu$-PIV method capable of 
submicrometric resolution \cite{Joseph2005}, the rheological setup 
lacks the sensitivity required to measure surface effects characterized 
by slippage in the micron range \cite{Boc06}. 
Our results
moreover points out that the effect of the underlying roughness can not be overlooked
in determining the slip length on such surfaces, therefore demonstrating that the gas cushion 
model used in Ref. \cite{Choi} fails to interprete slippage on nanopatterned surfaces.
\par
%
To conclude we have measured the surface hydrodynamic properties
of superhydrophobic surfaces made of carbon nanotube carpets incorporated into microchannels. 
Such surfaces are shown to exhibit a no slip boundary condition in the Wenzel state and slip lengths of a few microns in the superhydrophobic (Cassie) state. Moreover, these slip lengths were found to vary linearly with the lateral
roughness scale $L$ in agreement with theoretical predictions. For a
microsystem with given experimental constraints (working pressure,
dimensions), this now opens the possibility of designing optimized
surfaces in terms of friction reduction and robustness. 

We thank K. Stephan R. Ferrigno, S. Purcell and C. Barentin for their help during this work.

\end{document}